\documentclass[12pt]{amsart}
\usepackage[centertags]{amsmath}
\usepackage{graphicx}
\usepackage{amsfonts,amssymb,latexsym,epsfig,amsthm}

\usepackage{fullpage}

\newcommand{\ONE}{\pmb{1}}
\newcommand{\RR}{\mathbb{R}}

\newcommand{\XX}{\mathbb{X}}
\newcommand{\YY}{\mathbb{Y}}
\newcommand{\ZZ}{\mathbb{Z}}
\newcommand{\CC}{\mathbb{C}}
\newcommand{\Z}{{\mathbb{Z}^2_*}}

\newcommand{\Aone}{A_1^t(\mathcal{D},\gamma)}
\newcommand{\Atwo}{A_2^t(\mathcal{E},\eta)}
\newcommand{\Zp}{\Z}
\newcommand{\KoLong}{K_0(\gamma, \mathcal{E}, \eta,  \alpha')}
\newcommand{\Ko}{K_0}

\newcommand{\Pt}[1]{P_t(#1,\cdot)}
\newcommand{\Qt}[1]{Q_t(#1,\cdot)}
\newcommand{\bpf}[1][Proof]{{\noindent {\sc #1:    }}}
\newcommand{\epf}{{{\hspace{4 ex} $\Box$      \smallskip}}}
\newcommand{\ccdot}{ \ \cdot \ }

\newtheorem{theorem}{Theorem}
\newtheorem{lemma}{Lemma}[section]
\newtheorem{corolary}[lemma]{Corollary}
\newtheorem{proposition}[lemma]{Proposition}

\title[Small Scales of Stochastic Navier-Stokes Equations]{The Small
  Scales of the Stochastic Navier Stokes Equations under Rough Forcing}

\begin{document}
\author{Jonathan C. Mattingly  \and Toufic M.  Suidan}
\address{Jonathan C. Mattingly(jonm@math.duke.edu): School of Math, Institute for Advanced
  Study,  Princeton, NJ, USA and Department of Mathematics, Duke
  University,  Durham, NC 27708, USA. }
\address{Toufic Suidan: School of Math, Institute for Advanced Study,
  Princeton, NJ, USA and Courant Institute of Mathematical
  Sciences, New York University, NYC, NY, USA}

\begin{abstract}
  We prove that the small scale structures of the stochastically forced
  Navier-Stokes equations approach those of the naturally associated
  Ornstein-Uhlenbeck process as the scales get smaller.
  Precisely, we prove that the rescaled $k$-th spatial Fourier mode
  converges weakly on path space to an associated Ornstein-Uhlenbeck
  process as $|k|\rightarrow \infty$.

  In addition, we prove that the Navier-Stokes equations and the 
  naturally associated Ornstein-Uhlenbeck process induce equivalent
  transition densities if the viscosity is replaced with sufficient
  hyperviscosity. This gives a simple proof of unique ergodicity for
  the hyperviscous Navier-Stokes system. We show how different
  strengthened hyperviscosity produce varying levels of equivalence.

\end{abstract}
\maketitle

\section{Introduction}
We consider the stochastically forced Navier-Stokes equations
\begin{equation}
  \label{eq:vorticity}
  \left\{ \begin{aligned}
      &\frac{\partial \omega}{\partial t}(t,x)+  
      (u(x,t)\cdot \nabla)\omega(x,t) = \nu\Delta
      \omega(t,x) + \frac{\partial W}{\partial t}(t,x) \\
      &\omega(0,x)=\omega_0(x),  
    \end{aligned}\right. 
\end{equation}
on the two dimensional $2\pi$-periodic domain, $\mathbf{T}^2$. The
velocity, $u(x,t)$, is recovered from the vorticity, $\omega(x,t)$, by
the Biot-Savart law (see for instance \cite{b:MajdaBertozzi02}). We
assume that the fluid has no mean flow: $\int_{\mathbf{T}^2} u(x,t)dx
=0$. The stochastic forcing is generated by a Brownian motion of the
form $W(t,x)=\sum_{k \in \Z} \sigma_k \exp(i x \cdot k) \beta_k(t)$,
where $\Z$ denotes $\ZZ^2/\{0\}$, the $\sigma_k \in \CC$ are non-zero
complex coefficients satisfying $\sigma_k = \overline{\sigma_{-k}}$,
 and the $\beta_k$ are identically distributed standard complex Brownian
motions which are mutually independent except for the condition
$\beta_k=\overline{\beta_{-k}}$.  Setting $\omega(x,t)=\sum_{k \in \Z}
\omega_k(t) \exp(i k\cdot x)$, equation \eqref{eq:vorticity} becomes a
collection of coupled It\^o differential equations:
\begin{equation}\label{eq:SNSFourier}
  d\omega_k(t) =\Bigl[ -\nu |k|^{2}\omega_k(t) + 2i \sum_{j+\ell=k} \frac{\ell^\perp
  \cdot k}{|\ell|^2} \omega_\ell(t) \omega_j(t) \Bigr]dt + \sigma_k
  d\beta_k(t) \ .
\end{equation}
Some of our results will only hold for a modified version of
\eqref{eq:SNSFourier}  where the effect of the viscosity has been enhanced by
adding ``hyper-viscosity.'' For any $\alpha>0$, we consider the system
of equations
\begin{equation}
  \label{eq:SNS}
  d\omega_k(t) =\Bigl[ -\nu |k|^{\alpha}\omega_k(t) + 2i \sum_{j+\ell=k} \frac{\ell^\perp
  \cdot k}{|\ell|^2} \omega_\ell(t) \omega_j(t) \Bigr]dt + \sigma_k
  d\beta_k(t) \ .
\end{equation}
This is the Fourier representation of a partial
differential equation of the form \eqref{eq:vorticity}, where $\nu
\Delta$ has been replaced by $\nu \Delta^\frac\alpha2$.

Consider the Ornstein-Uhlenbeck process
\begin{equation}
  \label{eq:OUpde}
 \left\{ \begin{aligned}
  &\frac{\partial z}{\partial t}(t,x) = \nu\Delta^{\frac{\alpha}{2}}
  z(t,x) + \frac{\partial W}{\partial t}(t,x) , \\
  &z(0,x)=\omega_0(x).  
  \end{aligned}\right.  
\end{equation}
If $z(x,t)=\sum_{k \in \Z} z_k(t)\exp(i k\cdot x)$, then \eqref{eq:OUpde} becomes
\begin{equation}
  \label{eq:OU}
  dz_k(t)=-\nu|k|^{\alpha} z_k(t) dt + \sigma_k d\beta_k(t) \ .
\end{equation}
This Ornstein-Uhlenbeck process is the natural linear PDE associated
with the stochastic Navier-Stokes equations (SNS) \eqref{eq:SNS}.
Henceforth, we assume that for $k \neq 0$,
\begin{equation}
  \label{eq:decay}
  \frac{K_1}{|k|^l}\leq |\sigma_k| \leq  \frac{K_2}{|k|^l},
\end{equation}
for some positive constants $K_1$, $K_2$, and $l$. We assume
$\sigma_0=0$ in order to ensure that there is no mean flow.

In this note we give partial answers to the following questions.  When are the fine scale
structures of \eqref{eq:vorticity} the same as those of
\eqref{eq:OUpde} ?  Can $\omega_k$ be viewed as a
perturbation of $z_k$ when $|k|$ is large enough?  A lack of precise
understanding of the small scale structure is one of the major
technical impediments to a straightforward, Markovian analysis of many
stochastic partial differential equations. See
\cite{b:Mattingly03Pre} for a discussion of the
relationship between the small scale structures and some
approaches to proving ergodicity. 

The three theorems in this section offer answers to different aspects
of these questions. The first theorem demonstrates the weak
convergence of the small scales of (\ref{eq:vorticity}) to those of
(\ref{eq:OUpde}). The second theorem characterizes the relationship of
the small scales in terms of the equivalence of the Markov transition
densities. The third theorem characterizes this relationship in terms
of the equivalence on the entire path space of the dynamics.

Let $\omega_k'=\frac{\sqrt{2} |k|^\frac{\alpha}{2}}{|\sigma_k|} \omega_k$
and $z_k'=\frac{\sqrt{2} |k|^\frac{\alpha}{2}}{|\sigma_k|} z_k$. With
this rescaling, $z_k'$ is a complex Ornstein-Uhlenbeck process with
mean zero and variance one for any $k$.  First, we show that as
$|k_1|,\cdots,|k_d| \rightarrow \infty$,
$(\omega_{k_1}',\cdots,\omega_{k_d}')$ converges to
$(z_{k_1}',\cdots,z_{k_d}')$ on any finite time interval in a sense made
precise below. By $|k_1|,\cdots,|k_d| \rightarrow \infty$, we will always
mean $\displaystyle\min_{i\in\{1,\cdots,d\}} |k_i| \rightarrow \infty$.

\begin{theorem}\label{thm:week} In the above setting, the following
  two convergence results hold for any fixed finite $t>0$: 
  \begin{enumerate}
  \item Assume that the initial conditions satisfy: $|\omega_k(0)| <
    \frac{\mathcal{D}}{|k|^r}$ for some $\mathcal{D}, r>0$, such that
    $\displaystyle \limsup_{k\rightarrow \infty} |\sigma_k|^2
    |k|^{2r-\alpha} < 2$.  Then, for any bounded uniformly continuous 
    function $G:C([0,t];\CC^d)\rightarrow \RR$,
    \begin{equation}
      E|G(\omega_{k_1}',\cdots,\omega_{k_d}')-G(z_{k_1}',\cdots,z_{k_d}')|\rightarrow
      0, \nonumber  
    \end{equation} 
    as $|k_1|,\cdots,|k_d| \rightarrow \infty$, where $E$ denotes the
    expectation with respect to the driving Brownian motions. 
  \item For any continuous bounded function
    $G:C([0,t];\CC^d)\rightarrow \RR$,
    \begin{equation}
      E_{\mu^z}E|G(\omega_{k_1}',\cdots,\omega_{k_d}')-G(z_{k_1}',\cdots,z_{k_d}')|\rightarrow
      0, \nonumber
    \end{equation}
    as $|k_1|,\cdots,|k_d| \rightarrow \infty$.  Here, $E_{\mu^z}$
    denotes the expectation with respect to the initial conditions,
    the distribution of which is given by the stationary measure of
    the unscaled $z$ process.
  \end{enumerate}

\end{theorem}

While this is already an interesting statement, one might like strong
analytic control rather than weak convergence. In
\cite{b:FlMa95,b:Fe97}, the Bismut-Elworthy-Li formula was used to
prove the absolute continuity of the time $t$ transition densities of
the SNS starting from different initial conditions. This, in turn, was
used to prove a delicate ergodic theorem. In order to apply the
Bismut-Elworthy-Li formula, precise knowledge of the spatial
regularity of the SNS was needed. Their technique made use of the fact
that the $z(t)$ is less regular (in space) than $u(t)-z(t)$; hence,
the spatial regularity of $u(t)$ is determined by that of $z(t)$. In
light of this, one might hope to prove the stronger statement that the
distribution of $\{\omega_k\}_{k\in \Zp}$ is absolutely continuous
with respect to that of $\{z_k\}_{k\in \Zp}$.  One could think of this
holding either on path space or at a moment of time, $t$. For any
$\alpha>2$, we prove equivalence of the respective transition
densities. For $\alpha>4$, we prove equivalence on path
space.  Precisely, 
\begin{theorem}
  \label{thm:equivTimeT}
  For $\alpha > 2$ and $t>0$, the measures induced on
  $l^2(\Zp)$ by $z(t)$ and $\omega(t)$ are mutually absolutely
  continuous if $z(0)=\omega(0)$.
\end{theorem}
\begin{theorem}
  \label{thm:equiv}
  For $\alpha > 4$ and $t>0$, the measures induced on
  $C([0,t];l^2(\Zp))$ by $z$ and $\omega$ are mutually absolutely
  continuous if $z(0)=\omega(0)$.
\end{theorem}
Here, $l^2(\Zp)$ is the space of square summable sequences of complex
numbers indexed by $\Zp$.  In section \ref{sec:Uniqueness}, we use
this result to prove that the hyperviscous SNS system is uniquely
ergodic.  
\begin{corolary}\label{c:ergodic}
   If $\alpha >2$, then equation
  \eqref{eq:SNSFourier} has a unique invariant measure and this
  measure is equivalent to the unique invariant measure of equation
  \eqref{eq:OU}.  
\end{corolary}
This result could likely be proved using the methods in
\cite{b:FlMa95,b:Fe97}; in fact, it is weaker than the ergodic
results in these papers since it requires slight hyperviscosity.
However, they do not give the equivalence of the invariant measure of (\ref{eq:SNS})with respect to the invariant measure  of equation
\eqref{eq:OU}. More importantly, the method we present gives different
intuition about why the system is ergodic.  There are also methods to
prove equation \eqref{eq:vorticity} is ergodic by using estimates
which are fundamentally non-Markovian. See for example
\cite{b:EMattinglySinai00, b:BricmontKupiainenLefevere01,
  b:KuksinShirikyan00,b:Mattingly03Pre}.  The last reference contains
an overview of these less standard techniques.  Here, we stay in the
 Markovian framework. The analysis in this paper can be carried out for the Burgers equation without any hyperviscosity as the sums in the Girsanov term are one dimensional.

This paper is organized as follows. In section \ref{sec:deterministicEst}, we make several deterministic
observations about solutions to the SNS process and the associated
Ornstein-Uhlenbeck process.  In section \ref{sec:probEstimates}, we
estimate the probabilities of the deterministic picture suggested in
section \ref{sec:deterministicEst} and show that this picture is
correct with high probability.  In section \ref{sec:LimiteThm}, we
prove the small scale limit theorem. In section \ref{sec:Ergo}, we
prove the unique ergodicity of the hyperviscous Navier-Stokes
equations by proving that the SNS and the Ornstein-Uhlenbeck processes
are absolutely continuous on path space for $\alpha >4$ and have
absolutely continuous time $t$ marginals at any fixed time $t$ if $\alpha>2$.

\section{Deterministic Observations}
\label{sec:deterministicEst}
We define the following useful norms and subsets of path
space. Let $\omega=\{\omega_k\}_{k\in\Zp}\in{l^2(\Zp)}$.  Define
\begin{equation*}
  |\omega|_{\infty,\gamma}=\sup_{k\in\Zp} |k|^{\gamma}|\omega_k| \quad \mbox{
   and } \quad
   \|\omega\|=\left(\sum |\omega_k|^2\right)^\frac12,
\end{equation*}
and the subsets of path space
\begin{align*}
  \Aone&=\big\{z\in C([0,t],l^2(\Zp)): |z(s)|_{\infty,\gamma}\leq
  \mathcal{D}, \hspace{.1in} \forall s\in [0,t] \big\},\\
  \Atwo&=\big\{\omega\in C([0,t],l^2(\Zp)): \|\omega(0)\|^2 \leq \mathcal{E},
  \|\omega(s)\|^2 \leq \mathcal{\eta E}, \hspace{.1in} \forall s\in [0,t]
  \big\} .
\end{align*}
The arguments in this section are related to those in
\cite{b:MattinglySinai98}. We define ``trapping''
regions, along whose boundary the vector field corresponding to the dynamics points inward; hence,
solutions are trapped within these regions for all time.
 
Let $\alpha'\in (1,\alpha]$,  and define $\KoLong$ to be the smallest integer such
that
\begin{equation*} 
  \bigg{(}C(\gamma)\sqrt{\eta \mathcal{E}} 
  \frac{|k|\sqrt{\log{|k|}}}{\nu |k|^{\alpha'}}\bigg{)} < \frac{1}{6},
\end{equation*}
for all $k$ such that $|k|>\Ko$. $C(\gamma)$ is a constant which
only depends on $\gamma$ through summation formulas; it is made
explicit in the appendix.

It will be useful to set $\mathcal{D}^{\prime} =2\sqrt{\eta
  \mathcal{E}} \Ko^{\gamma}$. This constant is picked to ensure that the enstrophy,
$\sqrt{\eta \mathcal{E}}$, helps control some $|\cdot|_{\infty,
  \gamma}$ norm , once $\Ko$ is determined.

As stated before, we are interested in comparing solutions of the Ito
stochastic differential equations given in \eqref{eq:SNS} and
\eqref{eq:OU}. To accomplish this, we study the difference of
these two processes.  Let $\rho=\omega-z$ and $\rho(0)=0$.
$\rho=\{\rho_k\}_{k\in\Zp}$ satisfies the system of random ordinary differential equations:
\begin{equation}
\frac{d\rho_k(t)}{dt}=-\nu |k|^{\alpha}\rho_k(t) + F(\omega)_k(t)
\end{equation}
where $F(\omega)_k(t)$ is the nonlinear term in the drift of \eqref{eq:SNS}.
The following proposition gives sufficient control of $\rho_k(t)$.
\begin{proposition}\label{prop1}
  If $z\in\Aone$ and $\omega\in \Atwo$, then 
  \begin{align*}
    \sup_{s\in [0,t]} |\rho_k(s)|
  \leq \frac{2\overline{\mathcal{D}}}{|k|^{\gamma + (\alpha - \alpha')}}
  \end{align*}
  for all $k$ with $|k|> \Ko(\gamma,\mathcal{E}, \eta, \alpha')$, where
  $\overline{\mathcal{D}}=2\max\{\mathcal{D},
  \mathcal{D}^{\prime}\}$. Hence, $\omega
  \in A_1^t(3\bar{\mathcal{D}},\gamma) \cap  \Atwo$.
\end{proposition}
Before proving this proposition, we state the following technical lemma
proved in the appendix.
\begin{lemma}
\label{l:3sum}
  If $\omega\in \Aone \cap \Atwo$, then 
  \begin{align*}
    \sup_{s\in [0,t]} |F(\omega)_k(s)| \leq (C(\gamma)\sqrt{\eta
      \mathcal{E}} |k|\sqrt{\log{|k|}})
    \frac{\overline{\mathcal{D}}}{|k|^{\gamma}} .
  \end{align*}
\end{lemma}

\bpf[Proof of Proposition \ref{prop1}] We begin by noting that for $k$ with $|k|
\leq \KoLong$, the estimate $|\omega_k(s)| \leq
\frac{\mathcal{D}^{\prime}}{|k|^{\gamma}}$ holds for all $s\in [0,t]$;
this is because $\omega\in \Atwo$, and $\mathcal{D}^{\prime}$ has been chosen so
that this estimate holds.

Suppose that for some $s\in [0,t]$ and some $k$ with $|k|>\Ko$,
$|\rho_k(s)| =\frac{2\overline{\mathcal{D}}}{|k|^{\gamma + (\alpha -
    \alpha')}}$.  Suppose that for all $k^{\prime}$ with $|k^\prime|>
\Ko$ and $|k'|\neq k$, $|\rho_{k^{'}}(s)| \leq
\frac{2\overline{\mathcal{D}}}{|k^{\prime}|^{\gamma + (\alpha -
    \alpha')}}$. We show that the vector field points inward at this
point; hence, $\{\rho_k\}$ cannot violate the inequality in
Proposition 2.1. By assumption on $z$ at time $s$,
\begin{equation}
|\omega_k(s)| \leq \frac{\mathcal{D}}{|k|^{\gamma}} +
\frac{2\overline{\mathcal{D}}}{|k|^{\gamma +(\alpha -\alpha')}} \leq
\frac{3\overline{\mathcal{D}}}{|k|^{\gamma}}.
\end{equation}
By the Lemma 2.2,
\begin{equation}
|F(\omega)_k(s)| \leq \big(C(\gamma)\sqrt{\eta \mathcal{E}}
|k|\sqrt{\log{|k|}}\big) \frac{6\overline{\mathcal{D}}}{|k|^{\gamma}}.
\end{equation}
Computing
$|k|^{\alpha}|\rho_k(s)|=\frac{2 \overline{\mathcal{D}}}{|k|^{\gamma}}
|k|^{\alpha'}$,  and using the fact that $|k| > \Ko$, we have 
\begin{equation}
\nu |k|^{\alpha}|\rho_k(s)| > |F(\omega)_k(s)|. \nonumber
\end{equation} 
Multiplying the equation for $\rho_k$ by $\bar \rho_k$ produces
\begin{align*}
  \frac12 \frac{d \ }{ds} |\rho_k(s)|^2 &= - \nu |k|^\alpha |\rho_k(s)|^2 +
  F(\omega)_k(s)\bar \rho_k(s)\\
  & \leq\big( - \nu |k|^\alpha |\rho_k(s)| +  |F(\omega)_k(s)|   \big)|\rho_k(s)|
\end{align*}
This implies that $|\rho_k(s)|$ must decrease at time $s$, since the
vector field of the random ODE for $\rho_k$ points inward. \epf

\section{Probabilistic Estimates}
\label{sec:probEstimates}
In this section, we show that for certain
choices of their defining parameters, the events or sets in path space defined
in the previous section occur with high probability.  The following two lemmas give
conditions under which these events occur with high probability.

\begin{lemma}
  Fix a $\mathcal{E} > 0$ and finite $t>0$. For any SNS initial
  condition, $\omega(0)$, satisfying $\|\omega(0)\|^2 \leq
  \mathcal{E}$,
\begin{equation}
P\left(\omega\in \Atwo \right)\geq
  1-\exp\left \{ -\frac{\nu}{\sigma_{max}^2}[(\eta -1)\mathcal{E} -
  \mathcal{E}_1 t] \right \},
\end{equation} 
for $\eta$ sufficiently large,
  $\sigma_{max}=\max_{k\in \Zp} |\sigma_k|$, and
  $\mathcal{E}_1=\sum_{k\in \Zp} |\sigma_k|^2$.
\end{lemma}

\bpf The proof of an almost identical result can be found in
\cite{b:Mattingly02}. By It\^o's formula, we have
\begin{align*}
  \|\omega(t)\|^2 =& \|\omega(0)\|^2 + \mathcal{E}_1 t + 2 \int_0^t\sum_k
  \sigma_k \cdot \omega_k(s)dB_k(s) - 2 \nu  \int_0^t\sum_k|k|^{\alpha}
  |\omega_k(s)|^2 ds\\
  =&\|\omega(0)\|^2 + \mathcal{E}_1 t + N_t,
\end{align*}
where $N_t \leq M_t - \frac12 \frac{\nu}{\sigma^2_{max}} \langle M
\rangle_t$, $M_t= 2 \sum_k \int_0^t \sigma_k \cdot \omega_k(s)dB_k(s)$,
and $\langle M \rangle_t$ is the quadratic variation of the martingale $M_t$.
The standard exponential martingale estimate
for $L^2$-martingales gives
\begin{equation*}
P\bigg{(}\sup_{s\in[0,t]} N_s > (\eta -1)\mathcal{E} -\mathcal{E}_1
t\bigg{)} \leq \exp\bigg{\{}-\frac{\nu}{\sigma_{max}^2}[(\eta
-1)\mathcal{E} - \mathcal{E}_1 t] \bigg{\}}, 
\end{equation*}
for $\eta$ sufficiently large; this is the desired estimate. \epf 

We now state a simple lemma for the  Ornstein-Uhlenbeck process.
\begin{lemma}\label{l:trapAway}
  Fix $r>0$ so that $\displaystyle \limsup_{k\rightarrow \infty} |\sigma_k|^2
  |k|^{2r-\alpha} < 2$. If $|z(0)|_{\infty, r} \leq \infty$, then 
\begin{equation}
P\left (z\in A_1^t(\mathcal{D},r)\right )\rightarrow 1,
\end{equation} 
for any fixed $t>0$ as $\mathcal{D} \rightarrow \infty$.
\end{lemma} 

Under the conditions of the above lemmas: If $z(0)=\omega(0)$,
$|z(0)|_{\infty,\gamma} <\infty$,  and $\|\omega(0)\| \leq \mathcal{E}$, then
for any fixed $\delta >0$,  we can find $\mathcal{D}$ and $\eta$ so that
\begin{equation}  \label{wzProb}
  P\bigg{(} z\in A_1^t(\mathcal{D},r),\ \omega \in \Atwo \bigg{)} >
  1-\delta \ .
\end{equation}
Combining these lemmas with Proposition \ref{prop1}, we find that with
probability at least $1-\delta$,
  \begin{align*}
    \sup_{s\in
      [0,t]}|\rho_k(s)| \leq \frac{2\overline{\mathcal{D}}}{|k|^{r +
       ( \alpha - \alpha')}}, 
  \end{align*}
and 
\begin{equation*}
  P\bigg{(} \omega\in A_1^t(3\bar{\mathcal{D}},r) \cap \Atwo
  \bigg{)} > 1-\delta \ .
\end{equation*}
\section{Limit Theorem}
\label{sec:LimiteThm}
In this section, we prove Theorem \ref{thm:week} which states that the
high modes of the stochastic Navier-Stokes system, when scaled
appropriately, converge in $C[0,t]$ to a standard complex
Ornstein-Uhlenbeck process. We use the pathwise control gained in the
previous sections to prove this.

Let $G:C([0,t],\CC^d)\rightarrow \RR$ be a bounded uniformly
continuous function.  Fix $\epsilon >0$. Pick $\delta>0$ so that if
$\|x-y\|_\infty<\delta$, then $|G(x)-G(y)|<\epsilon$. Here,
$\|\ccdot\|_{\infty}$ denotes the sup norm in time.  Define $z_k'=
\frac{\sqrt{2} |k|^\frac{\alpha}{2}}{|\sigma_k|} z_k$,
$\omega_k'=\frac{\sqrt{2} |k|^\frac{\alpha}{2}}{|\sigma_k|} \omega_k$,
and $\rho_k'=\omega_k'-z_k'$.  Note that $z_k'$ is a standard complex
Ornstein-Uhlenbeck process with mean 0 and variance 1. Fix
$\epsilon'>0$. By the estimates of the previous two sections and the
assumption on the initial condition, we know that there exists a
constant, $K$, such that
$P(\|\rho_{k_1}'\|_\infty+\cdots+\|\rho_{k_d}'\|_\infty <
\delta)>1-\epsilon'$ for all $|k_i|\geq K$.  This gives
\begin{equation*}
  E\big|G(\omega_{k_1}',\cdots,\omega_{k_d}')-G(z_{k_1}',\cdots z_{k_d}')\big| \leq
  \epsilon P\big(
  \|\rho_{k_1}'\|_\infty+\cdots+\|\rho_{k_d}'\|_\infty < \delta\big) + 2 \overline{G}
  \epsilon'  
\end{equation*}
Here, $\overline{G}=\sup_x |G(x)|$. Since $\epsilon, \epsilon'$ are
arbitrary, we have proven the first part of Theorem \ref{thm:week}.

Now let $G:C([0,t];\CC^d)\rightarrow \RR$ be a bounded continuous function. Fix
$\epsilon>0$. Let $\delta_n$ be a sequence of positive numbers tending
to $0$ and define 
\begin{equation*}
A_n=\{x\in C([0,t];\CC^d): \|x-y\|_\infty<\delta_n
\Rightarrow |G(x)-G(y)|<\epsilon\}.
\end{equation*}  
Since $G$ is a continuous function, $\bigcup_{n=1}^\infty A_n=
C([0,t];\CC^d)$.  Setting $k_{*}=\min_i |k_i|$, the definition of
$A_n$ implies that
\begin{multline}\label{eq:EestimateWeak}
  E_{\mu^z} E
  \big|G(\omega_{k_1}',\cdots,\omega_{k_d}')-G(z_{k_1}',\cdots,z_{k_d}')\big|
  \leq 2 \overline{G} P\big((z_{k_1}',\cdots,z_{k_d}') \in
  A_{k_*}^c\big) \\ + \epsilon
  P\big(\|(\omega_{k_1}',\cdots,\omega_{k_d}')-
  (z_{k_1}',\cdots,z_{k_d}')\|_\infty < \delta_{k_*};
  (z_{k_1}',\cdots,z_{k_d}') \in A_{k_*}\big) \\ + 2
  \overline{G}P\big( (z_{k_1}',\cdots,z_{k_d}') \in
  A_{k_*}; \|(\omega_{k_1}',\cdots,\omega_{k_d}')-
  (z_{k_1}',\cdots,z_{k_d}')\|_\infty \geq \delta_{k_*}\big).
\end{multline}
Notice that
\begin{align*}
  \mu^z ( (z_{k_1}', \cdots, z_{k_d}') \in A_{k_*} ) \rightarrow 1 \mbox{
    as $|k_1|,\cdots, |k_d| \rightarrow \infty$}
\end{align*}
since for all $k$ the $z_k'$ are identically distributed and $ A_{k_*}
\rightarrow C([0,t];\CC^d)$ as $|k_1|,\cdots, |k_d| \rightarrow \infty$.
This estimate insures that the first term in \eqref{eq:EestimateWeak}
goes to zero as $|k_1|,\cdots, |k_d| \rightarrow \infty$.

By combining the estimate from \eqref{wzProb}
and Proposition \ref{prop1} at the end of the last section, we see
that if $\delta_n \rightarrow 0$ sufficiently slowly then
 \begin{equation*}
   P\big(\|(\omega_{k_1}',\cdots,\omega_{k_d}')-
  (z_{k_1}',\cdots,z_{k_d}')\|_\infty \geq 
  \delta_{k_*}\big) \rightarrow 0 \mbox{ as $|k_1|,\cdots, |k_d| \rightarrow \infty$.}
 \end{equation*}
 Thus, the third term in \eqref{eq:EestimateWeak} goes to zero.  Since
 the second term is bounded by $\epsilon$, which was arbitrary, we
 have proven the second statement of the theorem. \epf

 This shows that $(\omega_{k_1}',\cdots,\omega_{k_d}')$ approaches a
 standard d-dimensional complex Ornstein-Uhlenbeck process in
 distribution as $|k_1|,\cdots,|k_d| \rightarrow \infty$. We remark that
 for any fixed indices $k_1,\cdots,k_d$, Girsanov's theorem establishes
 equivalence of the measures on path space. But, the estimates on the
 Novikov term worsen as the indices tend to $\infty$. This is because
 the estimates on the nonlinearity (in the previous sections) grow in
 $|k|$.

\section{Ergodicity and Absolute Continuity}
\label{sec:Ergo}
\subsection{Absolute Continuity of $\omega$ and $z$ in Path Space
 when $\alpha>4$}
\label{sec:a4}
In this subsection, we show that the measures induced by $\omega$ and $z$ on
$C([0,t],l^2(\Zp))$ are equivalent (mutually absolutely continuous) if
$z(0)=\omega(0)$ and $\alpha>4$. We appeal to
Girsanov's theorem through Lemma \ref{l:compareMeasures}.

The equations governing $z(t)$ and $\omega(t)$ differ by the nonlinear
term. To apply Lemma  \ref{l:compareMeasures}, we need to show that
\begin{equation*}
 \int_0^t \sum_{k \in \Zp}
 \frac{|F_k(\omega(s))|^2}{|\sigma_k|^2} \ONE(\mathcal{B}) ds <
 C(\mathcal{B}) < \infty
\end{equation*}
for some measurable choice of $\mathcal{B} \subset C([0,t],l^2(\Zp))$ and some
constant $C$ which might depend on $\mathcal{B}$. Then, the Lemma 
implies that the measures on path space are equivalent when
restricted to $\mathcal{B}$. If for any $\delta>0$, one can find
such a $\mathcal{B}$, satisfying  $P\{\omega \in \mathcal{B}\} > 1- \delta$,
then Theorem \ref{thm:equiv} follows from  Lemma \ref{l:compareMeasures}.

Given any $\delta, \epsilon>0$, there exists a constant, $\mathcal{D}$, such
that 
\begin{equation}
P \left \{ z \in A_1^t\left (\mathcal{D}, \frac{\alpha}{2} +
    l-\epsilon\right) \right \} > 1-\delta.\nonumber 
\end{equation} 
Hence, by Proposition \ref{prop1}, there exist a constants,
$\mathcal{E}, \eta$, so that
\begin{equation}
P \left \{ \omega \in A_1^t(3\bar{\mathcal{D}}, \frac{\alpha}{2} +
  l-\epsilon) \cap \Atwo \right \} > 1- \delta. \nonumber  
\end{equation}
Set $\mathcal{B}= A_1^t(3\bar{\mathcal{D}}, \frac{\alpha}{2} + l
-\epsilon) \cap \Atwo$. For $\omega \in \mathcal{B}$, by Lemma \ref{l:3sum}, one has
\begin{align*}
  \sup_{s\in [0,t]} \sum_{k \in \Zp}
 \frac{|F_k(\omega(s))|^2}{|\sigma_k|^2}  \leq
 C(K_2,\bar{\mathcal{D}},\mathcal{E},\eta) \sum_{k \in \Zp}\frac{|k|^{2l}|k|^2 \log{|k|}
 }{|k|^{\alpha +2l -2\epsilon}} \ .
\end{align*}
Since $\epsilon$ is an arbitrary positive number, this sum is finite
if $\alpha >4$, proving the result.

\subsection{Absolute Continuity of Time $t$ Marginals of $\omega$ and $z$ when $\alpha>2$}

We show that if $\alpha >2$ and $t$ is fixed the distributions of
the $l^2(\Zp)$-valued random variables $\omega(t)$ and $z(t)$ are
mutually absolutely continuous. We use a technique from
\cite{b:HairerMattingly03Pre} which is inspired by a variation on the
Bismut-Elworthy-Li formula. In order to do this, we need Lemma \ref{l:relEnt} 
which controls convergence of densities given uniform control of associated relative entropies. Fixing the terminal time $t$, it is
sufficient to construct an auxiliary stochastic process
$\tilde \omega(s)$ such that $\omega(t)=\tilde \omega(t)$ and $\tilde
\omega$ is equivalent to $z$ on the path space $C([0,t],l^2(\Zp))$.

Setting
\begin{align*}
  \tilde F_k(s) = 
  \begin{cases}
    0 & s < \frac{t}2 \mbox{ or } s > t,\\
    2e^{-\nu|k|^\alpha (t-s)}F_k(\omega(2s-t)) & s \in [  \frac{t}2, t] 
  \end{cases}
\end{align*}
we define $\tilde w$ by
\begin{align*}
  d \tilde \omega_k(s)&= [ - \nu |k|^\alpha \tilde w_k(s)+ \tilde F_k(s)] ds + \sigma_k
  d\beta_k(s)\\
  \tilde \omega_k(0)&=\omega_k(0) \ .
\end{align*}
While $\tilde \omega(s)$ is not a diffusion, it is an adapted It\^o
process. Notice that 
\begin{align*}
  \tilde \omega_k(t)= e^{-\nu|k|^\alpha t}\omega_k(0)+ \int_0^t e^{-\nu|k|^\alpha
    (t-s)} \tilde F_k(s) ds + \int_0^t e^{-\nu|k|^\alpha
    (t-s)}  d\beta_k(s) \ . 
\end{align*}
The first and last term are identical to the first and last terms in the
analogous representation of $\omega_k(t)$. Observe that
\begin{align*}
  \int_0^t e^{-\nu|k|^\alpha (t-s)} \tilde F_k(s) ds &= \int_\frac{t}{2}^t
  e^{-\nu|k|^\alpha (t-s)} \tilde F_k(s) ds = \int_\frac{t}{2}^t
  e^{-\nu|k|^\alpha
    (t-s)}  2e^{-\nu|k|^\alpha (t-s)}F_k(\omega(2s-t)) ds \ .
\end{align*}
Setting $\tau=2s-t$, we have 
\begin{align*}
  \int_0^t e^{-\nu|k|^\alpha (t-s)} \tilde F_k(s) ds = \int_0^t
  e^{-\nu|k|^\alpha (t-\tau)}F_k(\omega(\tau)) d\tau .
\end{align*}
Hence, $\omega(t)=\tilde \omega(t)$. Observe that equality holds
only at time $t$ and that the distributions on path space are different.

We proceed to show that the auxiliary process, $\tilde \omega$,
induces a measure on the path space $C([0,t],l^2(\Z))$ which is
equivalent to the measure induced by the Ornstein-Uhlenbeck process.
This implies that the transition measures of the auxiliary process and
the Ornstein-Uhlenbeck process at time t are equivalent. Since the
transition measures of the hyperviscous Navier-Stokes equations and
the auxiliary process are equal at time t (by construction), we
conclude that the hyperviscous Navier-Stokes process ($\alpha>2$) and
the Ornstein-Uhlenbeck process have equivalent transition densities.
This fact leads to a simple proof of unique ergodicity for the
Navier-Stokes process; this proof is given in the next subsection.

We will first make precise the spaces in which we work. We let
\begin{equation*}
  (\Omega,\mathcal{F},\mathcal{F}_s, \mu)= (C([0,t], l^2(\Z)),
  \mathcal{F}, \mathcal{F}_s, P)  
\end{equation*}
where $\mathcal{F},\mathcal{F}_s$ are
the Borel sigma algebra and the filtration generated by finite
dimensional distributions up to time s, respectively. $P$ is the
measure induced on the path space by the Ornstein-Uhlenbeck process.
Note that one can recover the Brownian forcing from the Ornstein
Uhlenbeck process since all the relations are linear and invertible;
let $\tilde T:\Omega \rightarrow \Omega$ be the map that recovers the
Brownian paths from the Ornstein-Uhlenbeck process. Next, we let
$T:\Omega\rightarrow \Omega$ be the identity map.  Let $\tilde
\omega^{(N)}_k$ be defined for all $k\in \Z$ by:
\begin{equation*}
  d\tilde \omega^{(N)}_k(s)=[-|k|^\alpha \tilde \omega^{(N)}_k(s) +
  \tilde F_k(s) 1_{\{\tau_N>s\}}]ds+ \sigma_k d\beta_k(s). 
\end{equation*}
Here, $\tau_N=\inf\{s\in[0,t]: \|\omega\|>N, \mbox{ or } z\notin
A^s_1(N,\gamma)\}$ where $\gamma$ is fixed such that
$l+1<\gamma<l+\frac{\alpha}{2}$. $\tau_N$ is a stopping time and by
the earlier probabilistic estimates, $P$--almost surely
$\lim_{N\rightarrow \infty}\min\{\tau_N, t\}=t$. We note that $\tilde
F_k^{(N)}(s)=\tilde F_k(s) 1_{\{\tau_N>s\}}$ is a bounded Ito process.
Let $\tilde W_N:\Omega \rightarrow \Omega$ be the map which takes
Brownian paths in $\Omega$ to $\tilde \omega^{(N)}$. This is just the
solution map for the SDE for $\omega^{(N)}$. Let $T_N=\tilde W_N \circ
\tilde T$. Define $Q_N=T_N^*P$, the measure induced on $\Omega$ by
$\tilde \omega^{(N)}$. Since $\tau_N\rightarrow \infty$ as
$N\rightarrow\infty$ $P$--almost surely, for any $A\in \mathcal{F}$,
$Q_N(A)\rightarrow Q(A)$ where $Q$ is the measure induced on $\Omega$
by the process $\tilde \omega$.

Girsanov's theorem and a calculation imply that $P\sim Q_N$ for every
$N$. Before doing this calculation, we see that it implies that $Q\ll P$:
If $P(A)=0$ then
\begin{equation*}
Q(A)=\lim_{N\rightarrow \infty} Q_N(A)=\lim_{N\rightarrow \infty} 0 =0.
\end{equation*}
We have used the assumption that $Q_N(A)\rightarrow Q(A)$ for all
measurable $A$ and that $P\sim Q_N$. The calculation needed to prove
equivalence of the approximations and the Ornstein-Uhlenbeck process is: 
\begin{align*}
  \int_0^t \sum_{k \in \Zp} \frac{|\tilde
    F_k^{(N)}(s)|^2}{|\sigma_k|^2} ds \leq & 4
  \sum_{k \in \Zp} \left[\sup_{s \in[0,t]} |F_k(\omega(s))1_{\{\tau_N>s\}}|^2\right]\left[ \int_0^t
  \frac{e^{-2\nu|k|^\alpha (t-s)}}{|\sigma_k|^2} ds\right] \\ \leq & \sup_{s
    \in[0,t]} \sum_{k \in \Zp}\frac{2|F_k(\omega(s))1_{\{\tau_N>s\}}|^2}{\nu |k|^\alpha |\sigma_k|^2} 
  \leq poly(N)  \sum_{k \in
    \Zp}\frac{|k|^{2l}|k|^2 \log{|k|} }{|k|^{2\gamma + \alpha}}
   ,
\end{align*}
where $poly(N)$ is a fixed polynomial in $N$. This polynomial bound follows from Lemma 2.2 since $\tau_N>t$ implies that $\omega\in A_1^t(N,\gamma)\cap A_2^t(\|\omega_0\|,\frac{N}{\|\omega_0\|})$. Since $\gamma>l+1$ the last sum converges. Girsanov's theorem allows
us to assert that $P\sim Q_N$ for every $N$. (See Lemma
\ref{l:compareMeasures}  from the appendix.)

In order to show that $P\ll Q$, we will need a tail estimate on $P(\tau_N>t)$. We assume that $\mathcal{E}$ is fixed since it is just determined by
the initial condition.

\begin{equation*}
  P(z\notin A_1^t(\mathcal{D},\gamma)) \leq \sum_{k\in \Z} P\left (
    |z_k|>\frac{\mathcal{D}}{|k|^\gamma}\right ) 
\end{equation*}

A simple Gaussian tail estimate leads to
\begin{equation*}
  P\left ( |z_k|>\frac{\mathcal{D}}{|k|^\gamma}\right )\leq
  \frac{2|k|^{l+\frac{\alpha}{2} +\gamma}}{\mathcal{D} \sqrt{\pi}}
  e^{\frac{-\mathcal{D}^2}{2}|k|^{l+\frac{\alpha}{2}-\gamma}}
  e^{\frac{-\mathcal{D}^2}{2}|k|^{l+\frac{\alpha}{2}-\gamma}}. 
\end{equation*}
For $\mathcal{D}$ large enough but fixed,
$\frac{2|k|^{l+\frac{\alpha}{2} +\gamma}}{\mathcal{D} \sqrt{\pi}}
e^{\frac{-\mathcal{D}^2}{2}|k|^{l+\frac{\alpha}{2}-\gamma}}$ can be
made small uniformly in $|k|$.

It is an easy exercise to show that there is some fixed $C$ such that 
\begin{equation*}
  \sum_{k\in \Z}
  e^{\frac{-\mathcal{D}^2}{2}|k|^{l+\frac{\alpha}{2}-\gamma}}\leq C
  e^{-\frac{\mathcal{D}^2}{2}} 
\end{equation*}
for $\mathcal{D}$ sufficiently large.

Assume $\|\omega_0\|<\mathcal{E}$. By Lemma 3.1, we have that
\begin{equation*}
  P(\omega\notin \Atwo) \leq e^{\frac{-\nu}{\sigma_{max}^2}[(\eta
  -1)\mathcal{E} -\mathcal{E}_1t]} 
\end{equation*}
for $\eta$ sufficiently large. By the definition of $\tau_N$ and these exponential estimates, we see that there is a positive constant $c$ such that $P(\tau_N>t)< e^{-cN}$. We will use this bound to prove that $P\ll Q$. By the lemma A.2, it suffices to show that
$H(P|Q_N)$ is uniformly bounded: $\sup_N \int log(\frac{dP}{dQ_N})dP
<M<\infty$. Since the Radon-Nikodym derivative $\frac{dP}{dQ_N}$ is a local
exponential martingale, we need only show that:
\begin{equation*}
  \int  \left[\int_0^t \sum_{k \in \Zp} \frac{|\tilde
      F_k(s)|^2}{|\sigma_k|^2} ds\right]dP<\infty. 
\end{equation*}
In order to show this we apply Fatou's lemma and a simple stopping time argument. As usual, we denote by $E$ the expectation with respect to $P$.
\begin{eqnarray}
&&\int  \left[\int_0^t \sum_{k \in \Zp} \frac{|\tilde F_k(s)|^2}{|\sigma_k|^2} ds\right]dP \leq \lim_{N\rightarrow \infty} E\left [\sum_{k \in \Zp}\int_0^t \frac{|\tilde F_k^{(N)}(s)|^2}{|\sigma_k|^2} ds\right ] \nonumber \\
&&= \lim_{N\rightarrow \infty} \left \{ \sum_{k \in \Zp}\left ( E\left [ \int_0^t \frac{|\tilde F_k(s)|^2}{|\sigma_k|^2} ds 1_{\tau_N>t} \right ] + E\left[ \int_0^t \frac{|\tilde F_k^{(N)}(s)|^2}{|\sigma_k|^2} ds 1_{\tau_N \leq t } \right ] \right ) \right \} \nonumber\\
&&\leq  \lim_{N\rightarrow \infty} \sum_{k \in \Zp} E\left [ \int_0^t \frac{|\tilde F_k(s)|^2}{|\sigma_k|^2} ds 1_{\tau_N>t} \right ] + \lim_{N\rightarrow \infty} \left [ poly(N) e^{-cN} \sum_{k \in
    \Zp}\frac{|k|^{2l}|k|^2 \log{|k|} }{|k|^{2\gamma + \alpha}} \right ] \nonumber \\
&&= \lim_{N\rightarrow \infty} \sum_{k \in \Zp} E\left [ \int_0^t \frac{|\tilde F_k(s)|^2}{|\sigma_k|^2} ds 1_{\tau_N>t} \right ] = \lim_{N\rightarrow \infty} \sum_{l=1}^N \sum_{k \in \Zp} E\left [ \int_0^t \frac{|\tilde F_k(s)|^2}{|\sigma_k|^2} ds 1_{\tau_l>t\geq \tau_{l-1}} \right ]\nonumber \\ 
&&\leq  \left[\sum_{k \in \Zp}\frac{|k|^{2l}|k|^2 \log{|k|} }{|k|^{2\gamma + \alpha}}\right] \lim_{N\rightarrow \infty} \sum_{l=1}^N poly(l) e^{-c(l-1)}  < \infty.  \nonumber
\end{eqnarray}
This completes the proof of Theorem 2.

\subsection{Invariant Measures and Ergodicity}
\label{sec:Uniqueness}
In this section, we show that hyperviscous SNS has a unique invariant
measure, $\nu^\omega$. This measure is equivalent to the
Ornstein-Uhlenbeck invariant measure, $\nu^z$. By the preceding
section, we know that if $\alpha > 2$ then $\Pt{x} \sim \Qt{x}$, where
$\Pt{x}$, and $\Qt{x}$ are the transition kernels starting at $x$ for
the SNS process and Ornstein-Uhlenbeck process, respectively, and
$\sim$ denotes equivalence of measures.  Since $\Qt{x} \sim \Qt{y}$
for all $x,y \in l^2(\Zp)$ (simple to check since the semigroup of
Ornstein-Uhlenbeck is sufficiently contractive) $\Qt{y} \sim \Pt{x}$
for all $x,y \in l^2(\Zp)$. Invariance of the measures can be stated
as:
\begin{eqnarray*}
  \nu^\omega(A)= \int_{l^2(\Zp)} P_t(x,A) \nu^\omega(dx), \\
  \nu^z(A)= \int_{l^2(\Zp)} Q_t(x,A) \nu^z(dx).
\end{eqnarray*}
Existence of such a measure for the Ornstein-Uhlenbeck process is
immediate as it can be constructed explicitly. For the hyperviscous
SNS, existence follows from tightness arguments that have become standard
\cite{b:Fl94,b:ChowKhasminskii98}.

Suppose $\nu^\omega(A)>0$ and let $\epsilon_n=\frac{1}{n}$. Define
$B_n=\{x\in l^2(\Zp): Q_t(x,A)> \epsilon_n P_t(x,A)>0\}$. In order to
avoid any confusion, we remark that $P_t(x,A)=0$ if and only if
$Q_t(x,A)=0$ by our remarks on equivalence. Note that
\begin{equation*}
  \nu^z(A)= \int_{l^2(\Zp)} Q_t(x,A) \nu^z(dx) \geq \int_{B_n} Q_t(x,A)
  \nu^z(dx) \geq \epsilon_n \int_{B_n} P_t(x,A) \nu^z(dx). 
\end{equation*}
If $\nu^z(l^2(\Zp)/\bigcup_n B_n)=0$, then $\nu^z(A)>0$ since
$\nu^z(B_n)>0$ for some $n$ and $P_t(x,A)>0$ for all $x\in B_n$.  On
the other hand, let $\mathcal{L} =(l^2(\Zp)/\bigcup_n B_n)$ and
suppose $\nu^z(\mathcal{L})>0$. By the previous remark, for every
$x\in \mathcal{L}$, $Q_t(x,A)=0$; this implies $P_t(x,A)=0$. $\Pt{x} \sim
\Pt{y}$ for all $y\in l^2(\Zp)$, thus $P_t(y,A)=0$ for all $y\in
l^2(\Zp)$; but, this is impossible since $\nu^\omega(A)>0$. This
implies $\nu^z(B_n)>0$ for some $n$, so $\nu^\omega \ll
\nu^z$. Similarly, we can show $\nu^z \ll \nu^\omega$. Since $\nu^z
\sim \nu^\omega$ for any two $\omega$ invariant ergodic measures, we
know that these two measures are equivalent; therefore, they must be the same
measure by a standard ergodic theory argument. \epf

It is important to realize that not every infinite dimensional
Ornstein-Uhlenbeck process has transition densities which are
absolutely continuous for different initial conditions. This is true
in our setting because the semigroup is sufficiently 
contractive and the forcing decays slowly enough.

\section{Concluding Remarks}
\label{sec;conclusion} 

In this paper we have proven three theorems. They demonstrate
different ways to interpret the phrase ``the small scale are similar''. The results were given in increasing strength.  The
first is a weak convergence type of result. It states that the
rescaled modes of the Navier-Stokes equations converge to those of a
naturally associated Ornstein-Uhlenbeck process as the scales become
smaller or the wave number increase. The second theorem states that
the hyperviscous Navier-Stokes equations ($\alpha>2$) and its
associated Ornstein-Uhlenbeck process induce equivalent measure on
phase space at any fixed time $t$. In other words, the Markov
transition kernels  of the two processes at a fixed time are
equivalent.  This gives a simple proof of unique ergodicity for the
hyperviscous Navier-Stokes equations. The third theorem states that
the hyperviscous Navier-Stokes equations ($\alpha>4$) and its
associated Ornstein-Uhlenbeck process induce equivalent measure on
path space. As a result we see that the hyperviscous Navier Stokes equation has a unique invariant measure.

\section{Acknowledgments}
We would like to thank G{\'e}rard Ben Arous, Yuri Bakhtin, Martin
Hairer, \'Etienne Pardoux, and Yakov Sinai for
useful and informative discussions. We especially thank S.R.S. Varadhan for discussing the merits of relative entropy in proving convergence of approximate Girsanov densities. We also thank NSF for its support
through Grants DMS-9971087 in the case of the first author and
DMS-0202530 in the case of the second author. We also thank the
Institute for Advanced Study in Princeton for its hospitality and
support during the 2002-2003 academic year.
\appendix
\section{A Technical Lemma}
In this section, we prove the technical Lemma \ref{l:3sum}. It differs
little from the arguments of \cite{b:MattinglySinai98}. By
Cauchy-Schwartz,
\begin{align*}
  |F(\omega)_k(t)| \leq G(\omega)_k(t)=\sum_{l_1+ l_2 = k} |\omega_{l_1}(t)||\omega_{l_2}(t)| \big{|}
\frac{(k,l_2^{\perp})}{(l_2,l_2)} \big{|} \ .
\end{align*}
We estimate $\sup_{s\leq t} G(\omega)_k(s)$ given that $\omega\in \Aone \cap
\Atwo$. We begin by breaking the above sum into three parts:
\begin{equation*} 
  \Sigma_1= \sum_{|l_2| \leq \frac{|k|}{2}}, \hspace{.2in} \Sigma_2=
  \sum_{ 2|k| \geq |l_2| > \frac{|k|}{2}}, \hspace{.2in} \Sigma_3=
  \sum_{|l_2| > 2|k|}. 
\end{equation*}

1. To estimate $\Sigma_1$, we note that $\big{|}
\frac{(k,l_2^{\perp})}{(l_2,l_2)} \big{|} \leq \frac{|k|}{|l_2|}$ and
$|\omega_{l_1}| \leq \frac{2^{\gamma} \mathcal{D}}{|k|^{\gamma}}$. Hence
\begin{align*}
  \Sigma_1 \leq \frac{2^{\gamma} \mathcal{D}}{|k|^{\gamma-1}}
  \sum_{|l_2| \leq \frac{|k|}{2}} \frac{|\omega_{l_2}|}{|l_2|} &\leq
  \frac{2^{\gamma} \mathcal{D}}{|k|^{\gamma-1}} \sqrt{\sum_{l_2}
    |\omega_{l_2}|^2} \sqrt{\sum_{|l_2| \leq \frac{|k|}{2}}
    \frac{1}{|l_2|^2}} \\  
  &\leq \frac{2^{\gamma} \mathcal{D}}{|k|^{\gamma}} \sqrt{\eta
    \mathcal{E}} M |k| \sqrt{\ln{|k|}} 
\end{align*}
where $M$ is a constant arising from the second summation and does
not depend on any of the parameters.

2. To estimate $\Sigma_2$, we note that since $\frac{|k|}{2} < |l_2|
\leq 2|k|$, the inequalities $\big{|} \frac{(k,l_2^{\perp})}{(l_2,l_2)}
\big{|} \leq 2$ and $|\omega_{l_2}| \leq \frac{2^{\gamma}
  \mathcal{D}}{|k|^{\gamma}}$ hold. Thus,
\begin{equation*}
\Sigma_2 \leq \frac{2^{\gamma + 1} \mathcal{D}}{|k|^{\gamma}}
\sum_{|l_1|\leq 3|k|} |\omega_{l_1}| \leq \frac{2^{\gamma + 1}
  \mathcal{D}}{|k|^{\gamma}} \sqrt{\eta \mathcal{E}}
\sqrt{\sum_{|l_1|\leq 3|k|} 1} \leq \frac{2^{\gamma + 1}
  \mathcal{D}}{|k|^{\gamma}} \sqrt{\eta \mathcal{E}} (6|k|+1). 
\end{equation*}

3. Estimating $\Sigma_3$, we find
\begin{align*}
\Sigma_3 \leq |k| \sum_{|l_2|>2|k|} |\omega_{l_1}|\frac{|\omega_{l_2}|}{|l_2|}
&\leq |k| \sqrt{\eta \mathcal{E}} \sqrt{\sum_{|l_2|>2|k|}
  {\frac{|\omega_{l_2}|}{|l_2|}}^2} \\ 
&\leq |k| \sqrt{\eta \mathcal{E}} \mathcal{D} \sqrt{\sum_{|l_2|>2|k|}
  \frac{1}{|l_2|^{2(\gamma +1)}}} \leq |k| \sqrt{\eta \mathcal{E}}
\overline{M}(\gamma) \frac{\mathcal{D}}{{|k|^{\gamma}}}, 
\end{align*}
where $\overline{M}(\gamma)$ depends only on $\gamma$ through the
estimate on the last sum.

Adding the above estimates for the three sums, we see that
\begin{equation*}
\sup_{s \leq t}G(\omega)_k(s) \leq \sqrt{\eta \mathcal{E}}
\frac{\mathcal{D}}{|k|^{\gamma}} (|k|\sqrt{\ln{|k|}}) C(\gamma), 
\end{equation*}
which proves the lemma.

\section{Comparison of Measures on Path Space}
\label{sec:measurePath} 
Suppose that we have stochastic processes $X_i(t)$, $i=1,2$ on the
path space $C([0,T],\XX)$ where $\XX$ is some separable Hilbert space and $T\in
(0,\infty]$. Furthermore, assume that $X_i$ satisfies the equation
\begin{equation}
\begin{split}
dX_i(t)&=f_i(t,X_i[0,t])dt+gdW(t),\ t \in [0,T]\\
X_i(0)&=x_0.
\end{split}
\label{2SDDEs}
\end{equation}
For fixed $t$, the functions $f_1$ and $f_2$ map the space
$C_{[0,t]}=C([0,t],\XX)$ to $\XX$.  By $X[0,t]$ we mean the segment of
the trajectory on $[0,t]$. $W(t)$ is a cylindrical Brownian motion
over a separable Hilbert space $\YY$ and $g$ is a fixed invertible
Hilbert-Schmidt operator from $\YY \rightarrow \XX$.  For any
$\mathcal{B}\subset C_{[0,T]}$, define measures
$P^{(i)}_{[0,T]}(\ccdot ; \mathcal{B})$ on the path space as:
\begin{equation*}
  P^{(i)}_{[0,T]}(A; \mathcal{B}) = 
  P\{X_i[0,T]\in A\cap\mathcal{B}\},\ \mbox{\rm for}\ A \subset C_{[0,T]}.
\end{equation*}
Define also $D(t,\ccdot)=f_1(t,\ccdot)-f_2(t,\ccdot)$.

In this setting, we have the following result which is a variation on
Lemma B.1 from \cite{b:Mattingly02} and follows quickly from
Girsanov's Theorem.
\begin{lemma} \label{l:compareMeasures}  
  Assume there exists a constant $D_*\in(0,\infty)$ such that
  \begin{align}\label{eq:Novikov}
     \exp\left\{\frac12 \int_0^T \big|g^{-1}D\big(t,X_i[0,t]\big)\big|^2_{\YY} dt \right\}
    \ONE_\mathcal{B}(X_i[0,t])  < D_*
  \end{align}
   almost surely for $i=1,2$.
  Then the measures
  $P^{(1)}_{[0,T]}(\ccdot ; \mathcal{B})$ and $P^{(2)}_{[0,T]}(\ccdot ; \mathcal{B})$ are
  equivalent.
  
  \end{lemma}

  \bpf
  Define the auxiliary SDEs
  \begin{align*}
    dY_i(t) &=f_i\big(t, Y_i[0,t]\big) \ONE_{\mathcal{B}(t)}(Y_i[0,t]) dt + gdW(t),
  \end{align*}
  where 
  $
    \mathcal{B}(t)=\{ x \in C_{[0,t]}: \exists \bar{x} \in
    \mathcal{B} \mbox{ such that } x(s)=\bar{x}(s) \mbox{ for $s
    \in[0,t]$}  \} 
  $. 
  Solutions $Y_i(t)$ to these equations can be constructed as
  \begin{equation*}
    Y_i(t)=X_i(t)\ONE_{\{t\le\tau\}}+[gW(t)-gW(\tau)+X_i(\tau)]\ONE_{\{t>\tau\}}.  
  \end{equation*}
  Here $\tau=\inf\{s>0: X_i[0,s]\not\in \mathcal{B}(s)\}$.
  
  Denote
  $D_\mathcal{B}(t,x) =[f_1(t,x) - f_2(t,x)]\ONE_{\mathcal{B}(t)}(x)$.
  The assumption on $D$ in \eqref{eq:Novikov} and the definition of 
  $\mathcal{B}(t)$ imply that 
  \begin{equation*}
    \exp\left\{\frac12\int_0^T \big| g^{-1}D_\mathcal{B}
    \big(t,X[0,t]\big)\big|^2_{\YY} dt
    \right\}<D_*\quad \mbox{a.s.}
  \end{equation*}
  under both measures $P^{(i)}_{Y[0,t]}$ defining 
  solutions to auxiliary equation with $i=1$
  and $i=2$.
  Hence, Novikov's condition is satisfied for the difference in the drifts
  of the auxiliary equations. Girsanov's theorem implies that 
  $\frac{dP^{(1)}_{Y[0,t]}}{dP^{(2)}_{Y[0,t]}}(x) = \mathcal{E}(x)$, where
  the Radon--Nikodym derivative evaluated at a trajectory $x$ is defined by
  the stochastic exponent:
  \begin{align*}
    \mathcal{E}(x)=\exp\left\{\int_0^T \left\langle
        g^{-1}D_{\mathcal{B}}(s, x[0,s]), dW(s)\right\rangle_{\YY} -
      \frac12 \int_0^T |g^{-1}D_{\mathcal{B}}(s, x[0,s])|^2_{\YY} ds
    \right\}.
  \end{align*}
 
  Note that restrictions of measures $P^{(i)}_{Y[0,t]}$ on the set
  $\mathcal{B}$ coincide with $P^{(i)}_{[0,t]}(\ccdot;\mathcal{B})$.
  This proves that $P^{(1)}_{[0,t]}(\ccdot,\mathcal{B})$ is absolutely
  continuous with respect to $P^{(2)}_{[0,t]}(\ccdot;\mathcal{B})$.
  The reverse relation follows by symmetry and the proof is complete.
  \epf

\section{Relative Entropy and Equivalence of Measures}

The following lemma provides a sufficient condition for showing the
absolute continuity of a fixed measure with respect to measure arising
as the limit of certain approximating measures.

\begin{lemma}\label{l:relEnt}
  Let $(\Omega,\mathcal{F},\mu)$ be a probability space and let
  $(W,\mathcal{W})$ be a measure space. Assume $W$ is a Polish space
  and $\mathcal{W}$ is the Borel sigma algebra. Let
  $T:\Omega\rightarrow W$ and $T_n:\Omega\rightarrow W$ ($n=1,2,\cdots$)
  be measurable transformations. Let $P=T^*\mu$ and $Q_n=T_n^*\mu$ be
  the push--forward measures on $W$ induced by the respective
  transformations. Assume that there is a probability measure $Q$ on
  $W$ such that for any measurable $A\in\mathcal{W}$,
  $Q_n(A)\rightarrow Q(A)$. If $P \sim Q_n$ and $\limsup_{n\rightarrow
    \infty} |\int\frac{dP}{dQ_n} log \frac{dP}{dQ_n} dQ_n| <M<\infty$,
  then $P\ll Q$.
\end{lemma}  

\bpf Denote by $H(\mu|\nu)=\int \frac{d\mu}{d\nu}log\frac{d\mu}{d\nu}
d\nu$ the relative entropy of the probability measure $\mu$ with
respect to $\nu$ (when it exists). We begin by proving a basic
inequality. If $\mu$ and $\nu$ are mutually absolutely continuous,
$f\in L^1(\mu)$ and $H(\mu|\nu)<\infty$, then
\begin{equation*}
  \int f d\mu\leq H(\mu|\nu) + \log \left ( \int e^f d\nu\right ).
\end{equation*} 
This inequality follows from the simple calculation:
\begin{eqnarray*}
  \int f d\mu -\log \left ( \int e^f dv \right ) &=& \int f d\mu - \log
  \left ( \int e^f \frac{d\nu}{d\mu} d\mu \right ) \nonumber \\ &\leq&
  \int f d\mu - \int \log \left( e^f \frac{d\nu}{d\mu}\right ) d\mu =
  \int log \frac{d\mu}{d\nu} d\mu = H(\mu|\nu)\nonumber 
\end{eqnarray*}
In particular, for any $c>0$ the inequality becomes
\begin{equation*}
\int f d\mu\leq \frac{1}{c}H(\mu|\nu) + \frac{1}{c}\log \left ( \int
  e^{cf} d\nu\right ). 
\end{equation*}
Letting $f=\chi_A$, the characteristic function of a set
$A\in\mathcal{W}$, this inequality becomes
\begin{equation*}
P(A)\leq  \frac{1}{c}H(P|Q_n) + \frac{1}{c}log((e^c-1)Q_n(A) +1)
\end{equation*}
Fix $c>0$. If $Q(A)=0$, then $Q_n(A)\rightarrow 0$ by assumption.
Since $\limsup H(P|Q_n)<M<\infty$, as $n\rightarrow \infty$ the right
hand side is bounded by $\frac{2M}{c}$. $P(A)=0$ since $c$ is
arbitrary. Thus, $P\ll Q$. \epf


\begin{thebibliography}{EMS01}

\bibitem[BKL01]{b:BricmontKupiainenLefevere01}
J.~Bricmont, A.~Kupiainen, and R.~Lefevere.
\newblock Ergodicity of the 2{D} {N}avier-{S}tokes equations with random
  forcing.
\newblock {\em Comm. Math. Phys.}, 224(1):65--81, 2001.
\newblock Dedicated to Joel L. Lebowitz.

\bibitem[CK97]{b:ChowKhasminskii98}
Pao-Liu Chow and Rafail~Z. Khasminskii.
\newblock Stationary solutions of nonlinear stochastic evolution equations.
\newblock {\em Stochastic Anal. Appl.}, 15(5):671--699, 1997.

\bibitem[EMS01]{b:EMattinglySinai00}
Weinan E, J.~C. Mattingly, and Ya~G. Sinai.
\newblock Gibbsian dynamics and ergodicity for the stochastic forced
  {N}avier-{S}tokes equation.
\newblock {\em Comm. Math. Phys.}, 224(1), 2001.

\bibitem[Fer97]{b:Fe97}
Benedetta Ferrario.
\newblock Ergodic results for stochastic {N}avier-{S}tokes equation.
\newblock {\em Stochastics and Stochastics Reports}, 60(3--4):271--288, 1997.

\bibitem[Fla94]{b:Fl94}
Franco Flandoli.
\newblock Dissipativity and invariant measures for stochastic {N}avier-{S}tokes
  equations.
\newblock {\em NoDEA}, 1:403--426, 1994.

\bibitem[FM95]{b:FlMa95}
Franco Flandoli and B.~Maslowski.
\newblock Ergodicity of the {2-D} {N}avier-{S}tokes equation under random
  perturbations.
\newblock {\em Comm. in Math. Phys.}, 171:119--141, 1995.

\bibitem[HM]{b:HairerMattingly03Pre}
Martin Hairer and Jonathan Mattingly.
\newblock In preparation.
\newblock 2003.

\bibitem[KS00]{b:KuksinShirikyan00}
Sergei Kuksin and Armen Shirikyan.
\newblock Stochastic dissipative {P}{D}{E}s and {G}ibbs measures.
\newblock {\em Comm. Math. Phys.}, 213(2):291--330, 2000.

\bibitem[Mat02]{b:Mattingly02}
Jonathan~C. Mattingly.
\newblock Exponential convergence for the stochastically forced
  {N}avier-{S}tokes equations and other partially dissipative dynamics.
\newblock {\em Comm. Math. Phys.}, 230(3):421--462, 2002.

\bibitem[Mat03]{b:Mattingly03Pre}
Jonathan~C. Mattingly.
\newblock On recent progress for the stochastic {N}avier {S}tokes equations.
\newblock In {\em Journ\'ees \'Equations aux d\'eriv\'ees partielles},
  Forges-les-Eaux, 2003.
\newblock see
  http://www.math.sciences.univ-nantes.fr/edpa/2003/html/.

\bibitem[MB02]{b:MajdaBertozzi02}
Andrew~J. Majda and Andrea~L. Bertozzi.
\newblock {\em Vorticity and incompressible flow}, volume~27 of {\em Cambridge
  Texts in Applied Mathematics}.
\newblock Cambridge University Press, Cambridge, 2002.

\bibitem[MS99]{b:MattinglySinai98}
J.~C. Mattingly and Ya.~G. Sinai.
\newblock An elementary proof of the existence and uniqueness theorem for the
  {N}avier-{S}tokes equations.
\newblock {\em Commun. Contemp. Math.}, 1(4):497--516, 1999.

\end{thebibliography}
\end{document}